# Sunward propagating whistler waves collocated with localized magnetic field holes in the solar wind: Parker Solar Probe observations at 35.7 R$_\odot$ radii

O.V. Agapitov[1,2], T. Dudok de Wit[3], F.S. Mozer[2], J. W. Bonnell[2], J. F. Drake[4], D. Malaspina[5,6], V. Krasnoselskikh[2,4], S. Bale[2,7,8,9], P. L. Whittlesey[2], A. W. Case[10], C. Chaston[2], C. Froment[3], K. Goetz[11], K. A. Goodrich[2], P. R. Harvey[2], J. C. Kasper[12], K. E. Korreck[10], D.E. Larson[2], R. Livi[2], R. J. MacDowall[10], M. Pulupa[2], C. Revillet[3], M. Stevens[10], J. R. Wygant[11].

ABSTRACT

Observations by the Parker Solar Probe mission of the solar wind at ~35.7 solar radii reveal the existence of whistler wave packets with frequencies below 0.1 $f_{ce}$ (20-80 Hz in the spacecraft frame). These waves often coincide with local minima of the magnetic field magnitude or with sudden deflections of the magnetic field that are called switchbacks. Their sunward propagation leads to a significant Doppler frequency downshift from 200-300 Hz to 20-80 Hz (from 0.2 $f_{ce}$ to 0.5 $f_{ce}$). The polarization of these waves varies from quasi-parallel to significantly oblique with wave normal angles that are close to the resonance cone. Their peak amplitude can be as large as 2 to 4 nT. Such values represent approximately 10% of the background magnetic field, which is considerably more than what is observed at 1 a.u. Recent numerical studies show that such waves may potentially play a key role in breaking the heat flux and scattering the Strahl population of suprathermal electrons into a halo population.

[1] Corresponding author agapitov@ssl.berkeley.edu
[2] Space Sciences Laboratory, University of California, Berkeley, CA 94720
[3] LPC2E/CNRS-University of Orléans, Orléans, France
[4] University of Maryland, College Park, MD, USA
[5] Laboratory for Atmospheric and Space Physics, University of Colorado, Boulder, CO, USA
[6] Astrophysical and Planetary Sciences Department, University of Colorado, Boulder, CO, USA
[7] Physics Department, University of California, Berkeley, CA, USA
[8] The Blackett Laboratory, Imperial College London, London, UK
[9] School of Physics and Astronomy, Queen Mary University of London, London, UK
[10] Harvard-Smithsonian Center for Astrophysics, Cambridge, MA, USA
[11] University of Minnesota, Minneapolis, MN, USA
[12] University of Michigan, Ann Arbor, MI, USA

1. INTRODUCTION

In November 2018 Parker Solar Probe (PSP) became the first satellite mission to penetrate deep into the inner heliosphere, getting as close as 35.7 solar radii from the Sun. Between 2018 and 2024 this distance will progressively shrink to 9.8 solar radii ($R_\odot$), offering unique opportunities to study in situ the young solar wind (Fox et al. 2016). The mission addresses two fundamental problems in space physics: coronal plasma heating and the acceleration of solar wind plasmas. In both problems wave-particle interactions involving MHD and kinetic-scale waves (including whistlers) are known to play an important role.

During its first solar encounter PSP was nearly co-rotating with the Sun for more than one week and was immersed in a slow but highly alfvénic solar wind emerging from a small equatorial coronal hole (Kasper et al. 2019; Bale et al. 2019; Badman et al. 2020). As expected, in this type of solar wind the electron density and temperature increase with decreasing heliocentric distance while the electron $\beta$e - the ratio of electron thermal pressure to magnetic pressure - drops (Halekas et al. 2020). The Strahl becomes narrower and dominates the suprathermal fraction of the distribution. Halekas et al. (2020) report very low halo fractional densities near perihelion, much smaller than at larger heliocentric distances (McComas et al. 1992), smaller even than those previously reported at 0.3 a.u. (Maksimovic et al. 2005; Štverák et al. 2009). The electron halo and Strahl evolve with increasing radial distance from the Sun, with the fraction of the distribution in the halo increasing, and the fraction of the distribution in the Strahl decreasing (Maksimovic et al. 2005; Štverák et al. 2009). These changes presumably are the result of wave-particle interactions on the electron distribution, which may transform the Strahl into the halo through scattering by wave-particle interaction processes. Wave perturbations are observed by PSP continuously in solar wind in the MHD frequency range (Chaston et al. 2020; Krasnoselskikh et al. 2020; Mozer et al. 2020a) and at higher frequencies (Mozer et al. 2020b, Malaspina et al. 2020). Malaspina et al. (2020) showed that higher frequency plasma wave power enhancements manifest themselves in predominantly electric field fluctuations near 0.7 $f_{ce}$ and near 1.0 $f_{ce}$ with harmonics extending above $f_{ce}$. These waves were preliminarily identified as electrostatic whistler-mode waves and electron Bernstein modes; their duration ranges from seconds to hours. Wave amplitudes significantly increase with decreasing distance to the Sun (Malaspina et al. 2020; Mozer et al. 2020b) suggesting that these waves play an important role in the evolution of electron populations in the near-Sun solar wind. Here we focus on electromagnetic waves in the 20-100 Hz frequency range that generally coincide with local perturbations of the magnetic field. As will be shown later, these are Doppler shifted whistler waves.

One of the striking observations made by PSP during the first and third solar encounters is the omnipresence of rapid deflections of a magnetic field direction that is otherwise mostly radial. These so-called switchbacks are associated with an enhanced radial bulk plasma velocity and strongly affect the dynamics of the magnetic field (Kasper et al. 2019; Bale et al. 2019; Krasnoselskikh et al. 2020; Mozer et al. 2020a; Dudok de Wit et al. 2020). Some lead to a complete reversal of the magnetic field, hence the name switchback. These deflections are observed during the first and second solar encounters, in slow but highly alfvénic winds. They occur on time scales of seconds to hours and they are likely to be generated deep inside the corona (Dudok de Wit et al. 2020). Some switchbacks are accompanied by a small drop (of a few percent) in the amplitude of the magnetic field (Krasnoselskikh et al. 2020). The boundaries of these structures are plasma discontinuities that often have a significant normal component with respect to the magnetic field (Krasnoselskikh et al. 2020). Interestingly, they are accompanied by enhanced levels of wave activity (Krasnoselskikh et al. 2020, Mozer et al. 2020).

Most of the waves that are observed near or during switchbacks belong to the MHD and whistler frequency ranges. However, low frequency waves (with frequencies of a few Hz in the spacecraft frame) have also been observed; they have been identified as surface waves on the plasma discontinuities (Krasnoselskikh et al. 2020) and presumably are generated by surface velocity shift instabilities (Mozer et al. 2020a). In the following we concentrate on waves that belong to the whistler frequency range, motivated by the major impact whistler mode fluctuations are known to have on energetic electrons. In the solar wind such waves affect the heat flux through the scattering of Strahl electrons (Kajdic et al. 2016) while in the Earth's magnetosphere they control the dynamics of the population of relativistic electrons (Horne, 2007; Thorne, 2010). Whistler waves in the solar wind have been studied in detail at 1 a.u. (Lacombe et al. 2014) and, more recently, down to 0.3 a.u. with HELIOS observations (Jagarlamudi, private communication). Two potential sources of whistler waves in the solar wind are wave-particle interactions through electromagnetic instabilities and wave-wave interactions (Saito and Gary, 2007). PSP provides us with a unique opportunity to study these waves much deeper in the inner heliosphere, in regions where, precisely, they may influence the electrons populations of the young solar wind.

## 2. PSP OBSERVATIONS OF WHISTLER WAVES

In the following, we investigate whistler waves by means of electric and magnetic field fluctuations. PSP measures magnetic fluctuations between DC and typically 30 Hz with the MAG Fluxgate Magnetometer, and above typically 10 Hz with the SCM Search-Coil Magnetometer. The electric field is measured by two pairs of electric field antennas (EFI). The outputs of SCM and EFI are sampled by the DFB Digital Fields Board, which delivers a large variety of data products (Malaspina et al. 2016). All these instruments belong to the FIELDS consortium and are described in detail in (Bale et al. 2016). In what follows, we concentrate on waveforms that are sampled at 292.97 Hz although spectral matrices are also available for probing higher frequencies. The proton density and velocity are derived from Faraday cup that is a part of the SWEAP consortium (Kasper et al. 2016). These particle data are sampled every 12 s.

The first perihelion pass of Parker Solar Probe (PSP) occurred on November 7, 2018 at a distance of 35.7 solar radii. During the 4-5 days that preceded and followed the perihelion the unperturbed magnetic field was directed mostly sunward with a magnitude of approximately 50 to 70 nT. The bulk velocity of the solar wind was in the range of 300-340 km/s. A typical switchback structure that occurred on 4 November 2018 is illustrated in Figure 1. The reversal is best evidenced by the sudden change in sign of the radial component of the magnetic field, which is shown in red in Figure 1. For this particular event, which has been analyzed in detail by Krasnoselskikh et al. (2020), the magnetic field inside the switchback temporarily decreases from 70 nT to less than 50 nT. This structure has extended boundaries that last for several seconds, see Fig. 1a. Notice that the dip in the magnetic field amplitude does not coincide with the deflection; it starts approximately 10 seconds before the leading edge of the switchback and ends approximately 15 seconds after the trailing edge. These transition periods are marked with shaded bands in Fig. 1.

Both the leading and trailing edges of the switchback are accompanied by a short but conspicuous dip in the amplitude of the magnetic field, which drops by 30 nT (leading edge) and by 13 nT (trailing edge); these dips last for a few seconds. Both edges also coincide with an enhancement of the proton density, which rises from approximately 300 cm-3 to 600 cm-3. Switchbacks are always accompanied by an increase in wave activity, which is well illustrated in Fig. 1d by magnetic field fluctuations recorded by the SCM search-coil. Figure 1e shows the corresponding dynamic spectrum, which reveals broadband wave activity.

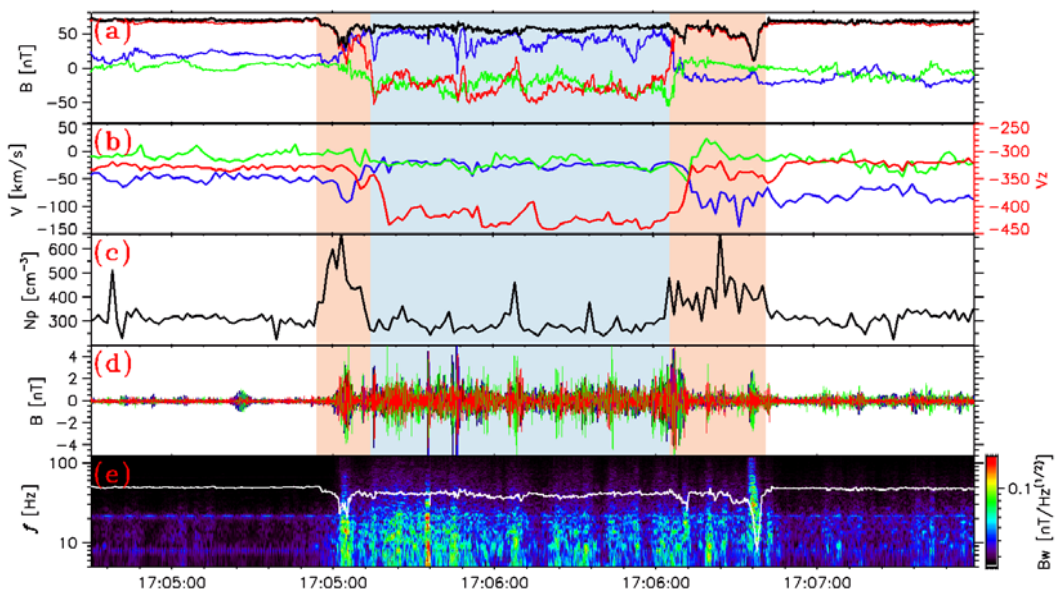

**Figure 1.** The magnetic field dynamics for a typical deflection (switchback) of the magnetic field observed during PSP's first solar encounter, on November 4, 2018, from 17:05 to 17:07 UT. The radial component of the magnetic field (red curve in panel (a)) exhibits an almost complete rotation inside the switchback and becomes negative (anti-sunward). The transverse components are shown in blue (x, in the ecliptic plane) and in green (y, transverse to the ecliptic plane). The magnitude is shown in black. Panel (b) represents plasma bulk velocity components (with a separate scale for the radial component $V_z$ shown in red) with the same color scheme as in panel (a). Panel (c) represents the proton density and Panel (d) the three components of magnetic field waveforms from SCM. The dynamic spectrum of these waveforms are shown in Panel (e), in which the solid white curve indicates the local lower hybrid frequency.

The local dip that occurs in the magnetic field at the extended leading edge of the switchback coincides with an enhancement of wave activity, see Figures 2a and 2b. The frequency of these waves is in the MHD range, below the local proton gyrofrequency whose Doppler-shifted frequency is between 1 and 3 Hz (Fig.2c); the corresponding frequencies in the plasma rest frame are 0.3-0.5 Hz. In that frequency range, the measured amplitude of the magnetic field reaches typically 10 nT, and the electric field 4 mV/m. The radial component of the Poynting flux in the plasma frame is negative, i.e. it is directed anti-sunward as is usual for waves that are observed in the solar wind. The ratio of magnetic to electric field wave power (Fig. 2e) agrees well with that of Alfvén waves with an effective antenna length of approximately 2.4 m (Mozer et al. 2020b).

The trailing edge of the switchback shown in Fig. 2 reveals a large-amplitude surface wave-like perturbation whose magnetic amplitude reaches 0.3-0.4 of the background field, see Fig. 2f. More details on the properties of these waves can be found in Krasnoselskikh et al. (2020). Notice in the local dip of the magnetic field a brief enhancement of higher frequency wave activity that is best seen in the electric field where it reaches amplitudes as large as 15 mV/m (see Fig. 2g) while in the magnetic field it goes up to 2 nT. In Fig. 3 we enlarge this small dip to highlight its coincidence with the wave packet, whose frequency ranges from 20 to 120 Hz, see Figs. 2h and 3. Interestingly, this wave propagates sunward as the radial component of the Poynting flux is significantly positive (Fig. 2i). Taking into account the Doppler shift the frequencies in the plasma frame should be considerably higher and belong to the whistler frequency range. This is confirmed by the value of the electric and magnetic field wave power ratio $E_w/B_w$ in Fig. 2j, which is significantly greater than expected from the dispersion relation for such low frequency waves.

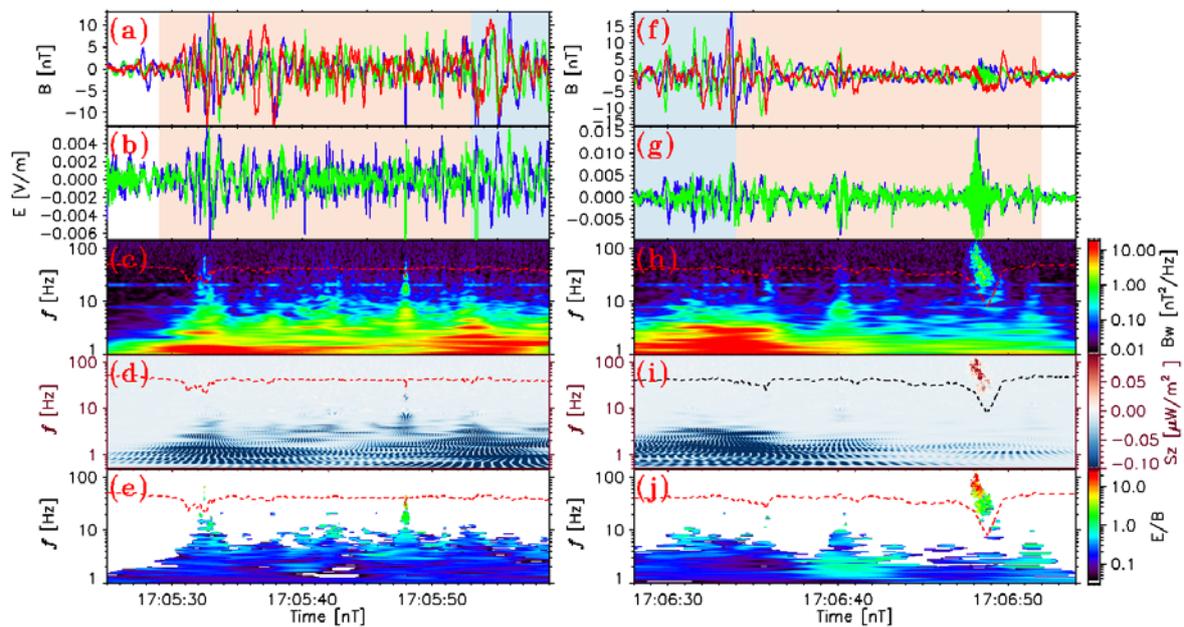

**Figure 2.** Enlargement of Fig. 1, showing magnetic and electric field fluctuations at the leading edge (left column) and trailing edge (right column) of the switchback. The red shaded time interval

corresponds to that shown in Fig. 1. Panels (a) and (f) show magnetic field fluctuations, and panel (b) and (g) show electric field fluctuations respectively, during the leading and trailing edges. The color scheme of the components is the same as in Fig. 1. The corresponding dynamic spectra of the magnetic field are given in panels (c) and (h). The signed dynamic spectra of the Poynting flux radial component are in panels (d) and (i). The ratio of wave power of electric and magnetic field perturbations is in panels (e) and (j).

In Figure 3 we zoom in the trailing edge of the same switchback and see that the local dip in the magnetic field is essentially caused by a decrease of its radial component. This dip coincides with an increase of the ratio between electron plasma frequency and electron gyrofrequency from 120 to approximately 500, see Fig. 3b. A polarization analysis reveals a right-handed circular polarization of the magnetic field and an elliptical polarization of the electric field with a $\pi/2$ phase shift. The dynamic spectrum in Fig. 3e shows a complex inner structure of the wave packet, which consists of a series of bursts. The phase shift of the magnetic and electric field components transverse to the radial direction attest a sunward propagation. Notice how the sign of the radial component of the Poynting vector (Fig. 3f) changes from positive (sunward) at high frequencies to negative (anti-sunward) at lower frequencies where, presumably, we have MHD waves. The frequencies of these wave packets fall between the lower hybrid frequency $f_{lh}$ (lower dashed curve in Figs. 3f and 3g) and one tenth of the electron cyclotron frequency $f_{ce}$ (upper dashed curve), similarly to what is known for whistler waves near 1 a.u. (e.g. Lacombe et al. (2014)). From all these properties we conclude that these are whistler wave packets.

The dispersion relation for cold plasma whistler waves gives us an E/B ratio that is significantly lower than the observed one, which appears highlighted in Fig. 3g. This suggests that the observed frequency range of our whistler waves is shifted down by the Doppler effect as the whistler phase velocity (300-500 km/s) is comparable to that of the plasma bulk velocity. To evaluate this Doppler shift and reconstruct the real wave frequency we need to evaluate the wave normal angle relative to the background magnetic field direction (shown in Fig. 3h) and the angle between the wave normal and the bulk velocity direction. The observed whistlers are found to have a wide range of wave normal angle values from quasi-parallel propagation to quasi-electrostatic propagating close to the resonance cone corresponding to the complex structure of the dynamics spectrum (Fig. 3b). Figure 3h thereby further supports the idea that our complex wave packet consists of a bunch of distinct and narrowband wave bursts.

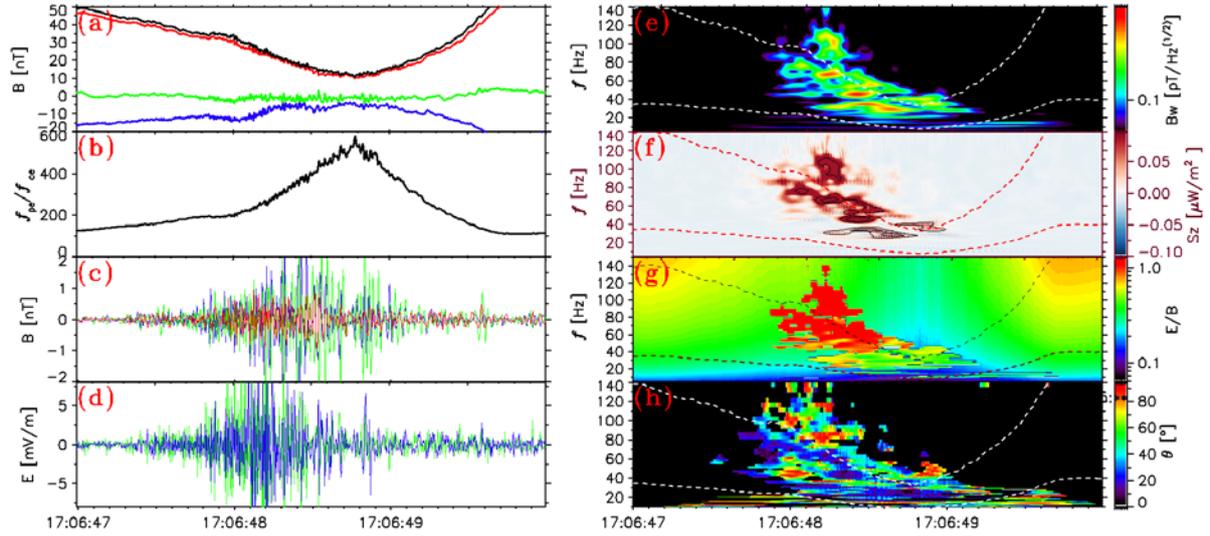

**Figure 3.** Enlargement of the trailing edge of the switchback of Fig. 1. Panel (a) shows the magnetic field from MAG with the same color code as in Figs.1 and 2. Panel (b) displays the ratio of electron plasma frequency $f_{pe}$ to electron gyrofrequency $f_{ce}$. Panels (c) and (d) show magnetic and electric field wave perturbations respectively. Panel (e) displays the dynamic spectrum of magnetic field perturbations $B_w$. The dashed curves in panels (e-h) represent the lower hybrid frequency (bottom curve) and 0.1 $f_{ce}$ (upper curve). Panel (f) displays the signed radial component of the Poynting flux. Red colors corresponds to a sunward propagation. Panel (g) displays the electric and magnetic field wave power ratio $E_w/B_w$: the background corresponds to the cold plasma approximation of the whistler wave dispersion relation while the highlighted area corresponds to observations. Panel (h) shows the wave normal angle relative to the direction of the background magnetic field.

We derive the wave frequency in the solar wind plasma frame from the Doppler shift and the whistler local parameters as obtained in the cold plasma approximation by making use of the wave normal angle values and the angle between wave normal and the bulk velocity direction. The reconstruction scheme is shown in Fig. 4a and the spectrum in the plasma frame is presented in Fig. 4b. The resulting frequencies of the wave packet are found to be in the range of 100-350 Hz, which corresponds to 0.2-0.5 of the local electron gyrofrequency $f_{ce}$. Wave normal angles (Fig.3h) vary for different whistler subpackets from close to parallel propagation to oblique (close to the resonance cone) that presumably reflect the effect of the propagation in inhomogeneous background magnetic field. The values of the $E_w/B_w$ wave power ratio estimated for whistlers with the resulting higher frequency ($0.2f_{ce}$-$0.5f_{ce}$) are sufficiently higher (~3-5 times) than the estimated for the observed frequency of 0.05-0.1 $f_{ce}$. While the observed whistler electric field (up to 10 mV/m) is closer to whistler dispersion parameters for the restored (to the plasma frame) wave frequency, we find that it is still ~3-4 times above the values that are estimated from the dispersion relation from the observed magnetic field power. This might be explained by the higher effective length of the electric field antennas (of typically 3.5-4.0 m) at higher frequencies.

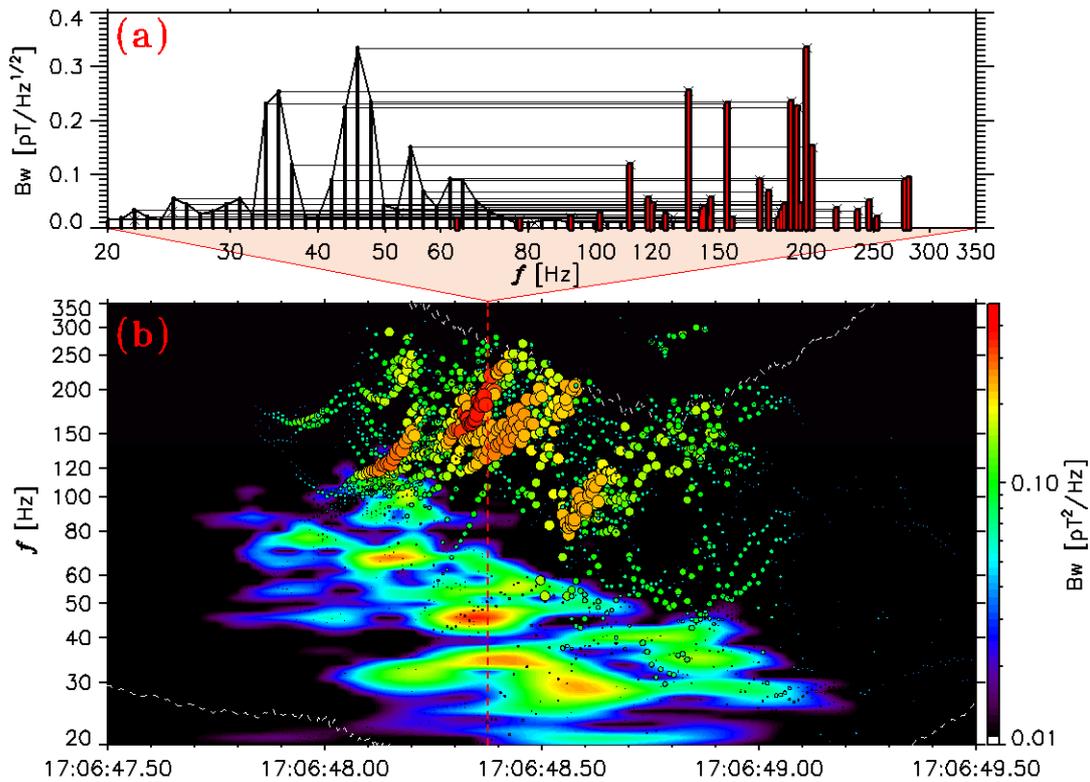

**Figure 4.** (a) - wavelet power spectrum of the whistler wave packet shown at 17:06:48.75 (the time moment is indicated by a red dashed vertical line in panel (b)). The spectrum estimated in the spacecraft frame is shown in black and the reconstructed spectrum in the solar wind frame (that takes into account the Doppler shift) is in red. The spectrogram in panel (b) compares the measured time-frequency dynamics spectrum estimated in the spacecraft frame with the reconstructed one, shown with circles (wave amplitude is color coded and indicated by the circles size).

The entire discussion has so far been based on one single whistler packet; two other examples will be given below. These events, however, are representative of the numerous ones that were observed during PSPs first solar encounter. More than 90% of them coincide with local depressions of the magnetic field or with sudden deflections of the magnetic field. The latter do not have to be complete switchbacks since partial deflections of the magnetic field also frequently give rise to whistler wave packets as long as the deflection is sudden and has the same characteristics as a complete switchback. The number of whistlers per day varies considerably and reflects the large variability of the number of switchbacks. During the first encounter we typically observe between 20 and 50 events per day that are unambiguously identified as whistler waves. This rate of occurrence is considerably larger than what has recently been predicted from HELIOS observations of whistlers at different distances from the Sun greater than 0.3 a.u. (Jagarlamudi, private communication)

## 3. DISCUSSION AND CONCLUSION

Let us now focus on the the properties of the sunward propagating whistler wave bursts as observed by Parker Solar Probe during the first solar encounter. The analysis shows that waves observed in the 20-100 Hz frequency range are electromagnetic right hand polarized whistlers propagating sunward both in the plasma frame and in the spacecraft frame; the value of their phase velocity value is usually higher than the bulk plasma velocity at 35.7 solar radii. These low-frequency electromagnetic whistler bursts are frequently associated with local minima of the background magnetic field magnitude. Two examples of whistler bursts (from the numerous cases captured on November 3-5 and having similar properties), both associated with local magnetic field magnitude minima, propagating sunward and captured on November 4, 2018 are presented in Fig. 5.

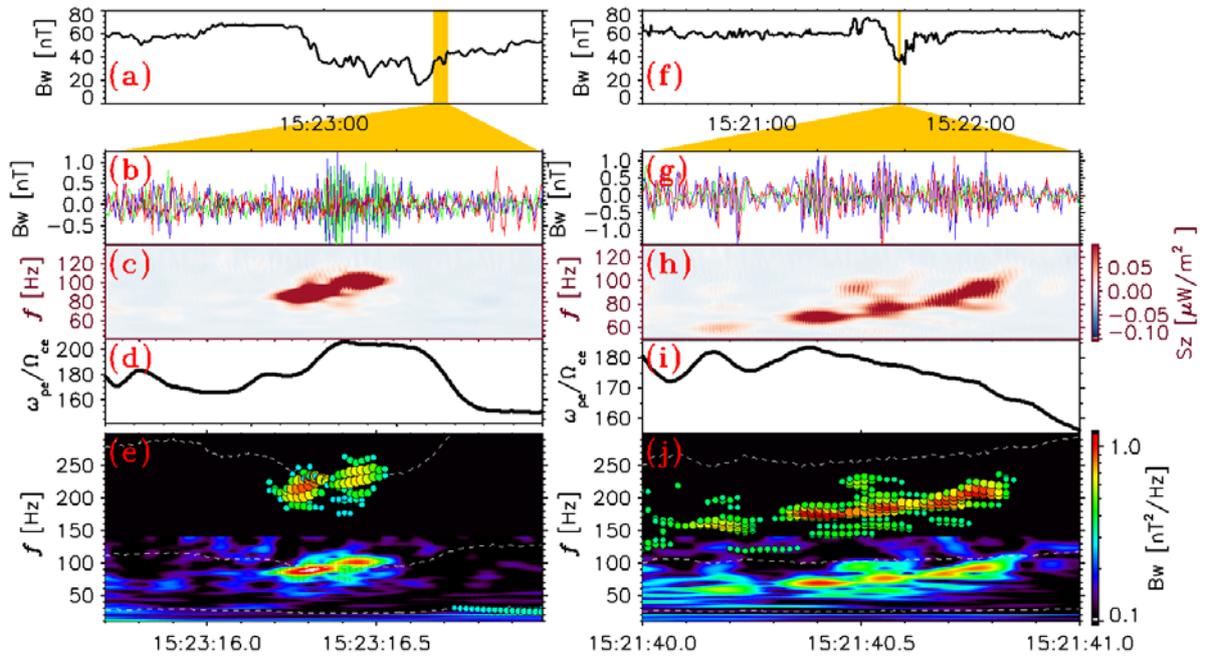

**Figure 5.** Whistler bursts in the local magnetic field minimums: (a) - the background magnetic field magnitude; (b) - waveforms of the wave magnetic field components; (c) the Poynting flux radial component (sunward direction is red); (d) the $f_{pe}/f_{ce}$ ratio; and (e) magnetic field dynamic wave power spectrum reconstructed to the plasma frame power spectrum is shown by the circles with color-coded by wave amplitude. Panels (f-j) represent a similar case.

The population of such sunward propagating whistlers can efficiently scatter the energetic particles of the solar wind and affect the Strahl population to spread their field-aligned pitch-angle distribution through pitch-angle scattering. The whistler resonance condition with electrons is given by $\omega - \vec{k}\vec{V} = \frac{n\Omega_{ce}}{\gamma}$ where $\vec{k}$ is wave vector; $\Omega_{ce} = 2\pi f_{ce}$; $\vec{V}$ is electron velocity, $n$ is an integer that can take on positive and negative values and $\gamma$ is the Lorentz factor. For sunward propagating whistlers with frequencies of 100-300 Hz, the resonance conditions are realized (due to inhomogeneities of the background magnetic field magnitude) for electrons with velocities between 3000 and 20000 km/s (~50 eV to 1 keV), which covers the observed Strahl energy range (Halekas et al. 2020). Such a scattering can be even more efficient when taking into account that a significant part of observed waves is oblique. Indeed, when the wave normal angles are found to be between the local Gendrin angle ($\cos\theta_G = 2f/f_{ce}$, Gendrin, 1961) and the local whistler resonance angle ($\cos\theta_{res} = f/f_{ce}$) effective scattering is strongly enhanced (Artemyev et al. 2013, 2014, 2016; Mourenas et al. 2013; Agapitov et al. 2014) on higher-order resonances. Such a scattering by high-amplitude whistler waves (whose amplitude reaches up to 10% of the background magnetic field magnitude) can regulate the heat flux as shown by (Roberg-Clark et al. 2019). For that reason the observed high-amplitude waves are likely to be an important factor in the dynamics of the solar wind distribution (Roberg-Clark et al. 2018a; Roberg-Clark et al. 2018b). The fraction of energetic electrons that belong to the halo distribution increases with the distance from the Sun while the fraction of Strahl population decreases (Maksimovic et al. 2005; Štverák et al. 2009; Halekas et al. 2020), which suggests a gradual transformation of the Strahl into the halo, presumably by pitch-angle scattering. Meanwhile, the angular width of the Strahl increases with radial distance (Hammond et al. 1996; Graham et al. 2017; Berčič et al. 2019). The whistler amplitudes that have been observed by PSP near the Sun are sufficiently larger than those observed 1 a.u. (Lacombe et al. 2014; Stansby et al. 2016; Tong et al. 2019a,b; Breneman et al. 2010). Their generation mechanism is presumably related to the cyclotron instability guided by a transverse temperature anisotropy of ~200 eV electrons and can be triggered by a magnitude gradient around the magnetic field magnitude minimum; this will be the subject for a future study (involving the electron distribution function processing). The statistical studies by Tong et al. (2019) showed a coincidence between the presence of whistlers and periods of higher temperature anisotropy. Numerical studies indicate electron beams as a possible source for whistler wave generation (Mourenas et al. 2017; Agapitov et al. 2015; Li et al. 2016; Kuzichev et al. 2019; Roberg Clark et al. 2019)

To conclude:

(1) PSP observations of electromagnetic whistler wave packets in the solar wind at ~35.7$R_\odot$ have revealed the existence of low-frequency (with frequencies of 20-80 Hz in the spacecraft frame, which is below 0.1 $f_{ce}$) whistler wave packets. These waves coincide with local minima of the magnetic field magnitude or with edges of magnetic switchbacks.

(2) These whistler waves are found to propagate sunward. Their phase velocity is in the range of 300-500 km/s, which leads to a significant Doppler frequency downshift from 200-300 Hz in the solar wind frame to 20-80 Hz in the spacecraft frame. This downshift allows these waves to be resolved by waveforms from magnetic (SCM and MAG) and electric (EFI) sensors, which are sampled at 292.97 Hz at perihelion.

(3) The whistler frequency in the plasma frame is of the order of 0.2-0.5 of the local electron gyrofrequency $f_{ce}$.

(4) The polarization of these waves varies in different wave packets from quasi-parallel to significantly oblique and close to the resonance values of wave normal angle (presumably due to the propagation in inhomogeneous background magnetic field).

(5) The wave amplitude reaches 2 to 4 nT, which corresponds to up to 10% of the background magnetic field. This amplitude is much larger than what is observed in the solar wind at 1 a.u. (Lacombe et al. 2014; Stansby et al. 2016; Tong et al. 2019a,b; Vasko et al. 2019). Such waves are very effective in scattering the Strahl population of solar wind electrons as shown recently in numerical simulations by Roberg-Clark et al. (2019). We conjecture that these whistler waves play a significant role in scattering the Strahl population and breaking the heat flux in the inner heliospheric solar wind.

ACKNOWLEDGEMENTS

OVA and JFD were supported by NASA grant 80NNSC19K0848; OA was partially supported by NSF grant number 1914670 and NASA Living with a Star (LWS) program (contract 80NSSC20K0218); TD, VK, CF and AL acknowledge the support of CNES and SDB the support of the Leverhulme Trust Visiting Professorship program. Parker Solar Probe was designed, built, and is now operated by the Johns Hopkins Applied Physics Laboratory as part of NASA's Living with a Star (LWS) program (contract NNN06AA01C). Support from the LWS management and technical team has played a critical role in the success of the Parker Solar Probe mission. All the data used in this work are available on the FIELDS data archive (http://fields.ssl.berkeley.edu/data/).